\documentclass[conference]{IEEEtran}
\IEEEoverridecommandlockouts
\usepackage{cite}
\usepackage{amsmath,amssymb,amsfonts}
\usepackage{algorithmic}
\usepackage{graphicx}
\usepackage{textcomp}
\usepackage[super]{nth}
\usepackage{xcolor}
\usepackage{nicefrac}
\def\BibTeX{{\rm B\kern-.05em{\sc i\kern-.025em b}\kern-.08em
    T\kern-.1667em\lower.7ex\hbox{E}\kern-.125emX}}
\begin{document}

\title{Feature Selection via Dynamic Graph--based Attention Block in MI--based EEG Signals
\footnote{{\thanks{This research was supported by the Challengeable Future Defense Technology Research and Development Program through the Agency For Defense Development (ADD) funded by the Defense Acquisition Program Administration (DAPA) in 2024 (No.912911601) was partly supported by the Institute of Information \& Communications Technology Planning \& Evaluation (IITP) grant, funded by the Korea government (MSIT) (No. RS-2019-II190079, Artificial Intelligence Graduate School Program (Korea University)).}
}}
}

\author{
\IEEEauthorblockN{Hyeon-Taek Han}
\IEEEauthorblockA{\textit{Dept. of Artificial Intelligence} \\
\textit{Korea University} \\ 
Seoul, Republic of Korea \\
ht\_han@korea.ac.kr}

\and

\IEEEauthorblockN{Dae-Hyeok Lee}
\IEEEauthorblockA{\textit{Dept. of Brain and Cognitive Engineering} \\
\textit{Korea University} \\
Seoul, Republic of Korea \\
lee\_dh@korea.ac.kr}

\and

\IEEEauthorblockN{Heon-Gyu Kwak}
\IEEEauthorblockA{\textit{Dept. of Artificial Intelligence} \\
\textit{Korea University} \\
Seoul, Republic of Korea \\
hg\_kwak@korea.ac.kr}
}

\maketitle

\begin{abstract}
Brain--computer interface (BCI) technology enables direct interaction between humans and computers by analyzing brain signals. Electroencephalogram (EEG) is one of the non--invasive tools used in BCI systems, providing high temporal resolution for real--time applications. However, EEG signals are often affected by a low signal--to--noise ratio, physiological artifacts, and individual variability, representing challenges in extracting distinct features. Also, motor imagery (MI)--based EEG signals could contain features with low correlation to MI characteristics, which might cause the weights of the deep model to become biased towards those features. To address these problems, we proposed the end--to--end deep preprocessing method that effectively enhances MI characteristics while attenuating features with low correlation to MI characteristics. The proposed method consisted of the temporal, spatial, graph, and similarity blocks to preprocess MI--based EEG signals, aiming to extract more discriminative features and improve the robustness. We evaluated the proposed method using the public dataset 2a of BCI Competition IV to compare the performances when integrating the proposed method into the conventional models, including the DeepConvNet, the M--ShallowConvNet, and the EEGNet. The experimental results showed that the proposed method could achieve the improved performances and lead to more clustered feature distributions of MI tasks. Hence, we demonstrated that our proposed method could enhance discriminative features related to MI characteristics.

\end{abstract}

\begin{IEEEkeywords}
brain--computer interface, electroencephalogram, motor imagery;
\end{IEEEkeywords}

\section{INTRODUCTION}
Brain--computer interface (BCI) is a system that facilitates direct interaction between humans and computers \cite{jeong2019classification, schalk2004bci2000}. This technology reclaims a novel way in neuroscience research by providing the approach to analyze brain signals associated with the activation of specific brain regions \cite{prabhakar2020framework}. In addition, electroencephalogram (EEG) is one of the non--invasive techniques for monitoring complex brain dynamics without the need for surgery \cite{nicolas2012brain, kim2015abstract}. EEG--based BCI systems have advantages including low--cost and high portability, which are promising for a wide range of applications such as a speller \cite{lee2018high}, a robotic arm \cite{zhou2023shared}, a wheelchair \cite{kim2018commanding}, or a drone \cite{nourmohammadi2018survey}.

The high temporal resolution of EEG signals enables real--time reflection of the human's cognitive state and intentions. This modality exhibit characteristics that are especially advantageous for various BCI paradigms, such as motor imagery (MI) \cite{bai2011prediction}. MI is a useful paradigm for controlling external electronic devices without physical movement \cite{cho2021neurograsp}. It is receiving more attention in the fields of rehabilitation and assistive technology \cite{wang2012translation}. However, EEG signals suffer from a low signal--to--noise ratio and various physiological artifacts, making it difficult to extract distinct features and analyze brain signals \cite{han2020classification}. Furthermore, EEG signals exhibit inherent variability, with more pronounced differences observed across individuals \cite{lee2020continuous}. 

To solve these problems, a large number of methodologies have been developed in recent years. Among them, the convolutional deep architectures are commonly employed as feature extractors to capture the significant features from MI tasks. Barmpas \textit{et al}. \cite{barmpas2023improving} proposed the dynamic convolutional approach based on causal reasoning to mitigate the inter--subject variability in decoding MI--based EEG signals, resulting in up to 5 \% improvement in generalization across subjects. Tang \textit{et al}. \cite{tang2023motor} focused on improving the classification performances of EEG signals in MI domain, which extracts various temporal and spatial features by the multi--scale convolutional framework, achieving the average accuracy of 96.87 \%. The convolution--based methods have shown impressive results in MI domain. However, EEG signals could contain features with low correlation to MI characteristics. Since the convolutional operation is limited to local receptive fields, a potential challenge arises in which weights of the deep model could become biased towards features with low correlation to MI characteristics. 

In this paper, we proposed the end--to--end deep preprocessing method that enhances the relevant MI characteristics while attenuating those with low correlation to MI characteristics. It was designed to be robustly integrated into the initial layers of the convolution--based methods. To extract the discriminative features for each MI characteristic, our proposed method was composed of the temporal and spatial convolutional layers. Additionally, we introduced the graph--based convolutional operation between two layers to consider the relationships across electrodes. By leveraging the implemented structure, we verified the effectiveness of the proposed method by integrating it with the baseline models. It was observed to be effectively preprocessed prior to being used as input for the baseline models. To the best of our knowledge, this is the first attempt to enhance discriminative features using dynamic graph--based attention block in MI--based EEG signals.

% %%%%%%%%%%%%%%%%%%%%%%%%%%%%%%%%%%%%%%%%%%%%%%%%%%%%%%%
\begin{figure}[t!]
\centerline{\includegraphics[width=\columnwidth]{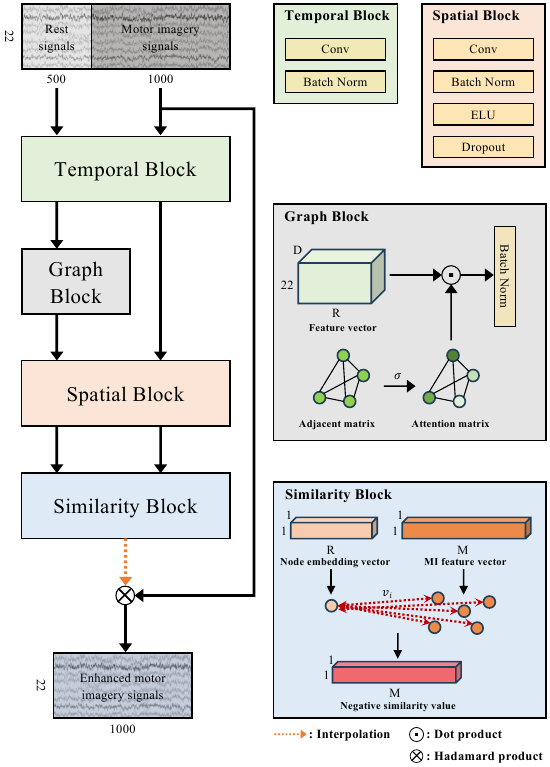}}
\caption{Visualization of the overall process in the proposed method. EEG signals of rest and MI are used as input simultaneously. The temporal, graph, and spatial blocks extract the significant features to effectively calculate the negative similarity. The similarity block contains a processing step that generates the value $\textit{v}_i$ to enhance MI features while attenuating rest features. ($D$: the dimension of the feature vector, $R$: the number of time steps compressed from rest signals, and $M$: the number of time steps compressed from MI signals).
}
\label{fig}
\end{figure}
% %%%%%%%%%%%%%%%%%%%%%%%%%%%%%%%%%%%%%%%%%%%%%%%%%%%%%%%

\section{MATERIALS AND METHODS}
\subsection{Dataset}
We utilized a dataset 2a of BCI Competition IV (BCIC2a) \cite{brunner2008bci}, which is the well--known public dataset commonly employed in the MI domain, to validate our experiments. BCIC2a consists of 22 electrodes EEG signals from nine subjects (A1--A9) at a sampling rate of 250 Hz. This dataset was collected from subjects performing four distinct tasks, including left hand (``L"), right hand (``R"), feet (``F"), and tongue (``T") with two sessions recorded on different days. Each session includes a total of 288 trials, with 72 trials per MI task. Specifically, we segmented the EEG data from 0 to 2 seconds as rest and from 2 to 6 seconds as MI signals.\\

\subsection{The Proposed Method}
To address the problem of bias towards features with low correlation to MI characteristics, we proposed the preprocessing method for signals. In this paper, the term of ``rest" is defined as the entire period excluding the period of MI tasks. The proposed method was designed to attenuate features that exhibit high similarity to rest characteristics in the entire temporal period of MI signals. Additionally, in the spatial aspect, the graph--based approach was introduced to learn the relationships between electrodes at different time points \cite {lee1992recognizing}. Fig. 1 showed the overall process of the proposed method. The proposed method was composed of four main components, including the temporal, spatial, graph, and similarity blocks. 

In the temporal and spatial blocks, each block employed a convolutional layer to extract significant features related to temporal and spatial information, respectively. Each kernel size was set to $1\times\nicefrac{\textit{F}}{32}$ in the temporal block and $\textit{C}\times1$ in the spatial block, where \textit{F} denotes the sampling rate and \textit{C} indicates the number of electrodes. The number of temporal and spatial kernels was set to 16. To prevent the weight bias problem during the feature extraction process, a batch normalization layer was applied after each convolutional layer. Additionally, we utilized the exponential linear unit (ELU) as an activation function and the dropout to avoid the overfitting problem after the convolutional layer in the spatial block. 

The graph block was designed between the temporal and spatial blocks to globally learn the relationships between electrodes from extracted temporal features. The graph--based convolutional operation was applied only to the rest features to more effectively attenuate features similar to rest characteristics in MI features. The graph--based convolutional operation is formulated as:
\begin{equation}
    \Tilde{A} = A+I,
\end{equation}
\begin{equation} 
    \Tilde{D} = \text{diag} \bigg (\bigg( \sum^C_{j=1} \Tilde{A}_{ij} \bigg)^{-\frac{1}{2}} \bigg),
\end{equation}
\begin{equation} 
    X' = \sigma(\Tilde{D}\Tilde{A}\Tilde{D}^T)X,
\end{equation}
where $\textit{A}\in\mathbb{R}^{\textit{C}\times\textit{C}}$ and $\textit{I}\in\mathbb{R}^{\textit{C}\times\textit{C}}$ are the adjacent matrix and the identity matrix, respectively. $\Tilde{\textit{A}}$ is the adjacent matrix with the added self--loop. $\Tilde{\textit{D}}\in\mathbb{R}^{\textit{C}\times\textit{C}}$ denotes the diagonal matrix representing the degree of $\Tilde{A}$, which is used to normalize the adjacent matrix. $\sigma$ is the sigmoid function as the activation function, and it is used to indicate electrodes with high correlation with rest characteristics as the attention. $\textit{X}'$ and $\textit{X}$ are the node embedding and feature vectors of rest signals, respectively.

In the similarity block, we utilized the negative cosine similarity method to find similar features to rest characteristics within MI signals. We averaged the node embedding vectors $\textit{X}'$ to determine the center vector, which was then used to calculate the negative similarity with the feature vectors of MI signals, where values closer to rest characteristics received lower points. We scaled the negative similarity points as [0, 1] to assign values to raw MI signals. Since we directly preprocessed raw EEG signals by our proposed method, MI signals with enhanced characteristics can be used as input regardless of the conventional models. It was employed to match the temporal length of the negative similarity vector to that of input (i.e., raw MI signals).\\

\subsection{Data Preprocessing and Evaluation Setting}
We utilized a $4^{th}$--order Butterworth bandpass filter between 0.5 and 38 Hz to capture the MI--relevant frequencies from BCIC2a. Additionally, the exponential moving standardization was utilized as the normalization method to reduce noise and fluctuations in BCIC2a. We trained three conventional models which are most commonly used in BCI paradigms, including the DeepConvNet \cite{schirrmeister2017deep}, the M--ShallowConvNet \cite{kim2022rethinking}, and the EEGNet \cite{lawhern2018eegnet} for 300 epochs. To evaluate the effectiveness of the proposed method, we compared the performances of three conventional models with (w/) and without (w/o) integration. Our configuration included a batch size of 64, a learning rate of 0.002, and a weight decay of 0.075. In addition, the AdamW optimizer and the CosineAnnealingLR scheduler were employed to improve training stability.\\

% %%%%%%%%%%%%%%%%%%%%%%%%%%%%%%%%%%%%%%%%%%%%%%%%%%%%%%%
\begin{table}[t!]
\caption{Comparison of the performances with the conventional models, including the DeepConvNet, the M--ShallowConvNet, and the EEGNet.}
\normalsize
\begin{center}
\renewcommand{\arraystretch}{1.4}{
\resizebox{\columnwidth}{!}{
\begin{tabular}{ccccccc}
\cline{2-7}
      & \multicolumn{2}{c}{DeepConvNet \cite{schirrmeister2017deep}} & \multicolumn{2}{c}{M--ShallowConvNet \cite{kim2022rethinking}} & \multicolumn{2}{c}{EEGNet \cite{lawhern2018eegnet}}     \\ \cline{2-7} 
      & avg. (\%) $\uparrow$             & std. $\downarrow$         & avg. (\%) $\uparrow$               & std. $\downarrow$            & avg. (\%) $\uparrow$            & std. $\downarrow$         \\ \hline
w/o   & 59.49           & 7.37          & 64.89             & 11.39            & \textbf{69.19} & \textbf{6.07} \\ 
w/    & \textbf{62.35}  & \textbf{7.35} & \textbf{68.67}    & \textbf{11.07}   & 69.08          & 6.24          \\ \hline
Diff. & 2.86            & --0.02         & 3.78              & --0.32            & --0.11          & 0.17          \\ \hline
\end{tabular}}}
\label{tab1}
\end{center}
\footnotesize{{$^*$avg.: average accuracy, std.: standard deviation, Diff.: difference, \\ w/o: without the proposed method, w/: with the proposed method}}
\end{table}
% %%%%%%%%%%%%%%%%%%%%%%%%%%%%%%%%%%%%%%%%%%%%%%%%%%%%%%%

% %%%%%%%%%%%%%%%%%%%%%%%%%%%%%%%%%%%%%%%%%%%%%%%%%%%%%%%
\begin{figure}[t!]
\centerline{\includegraphics[width=0.93\columnwidth, height=0.53\textwidth]{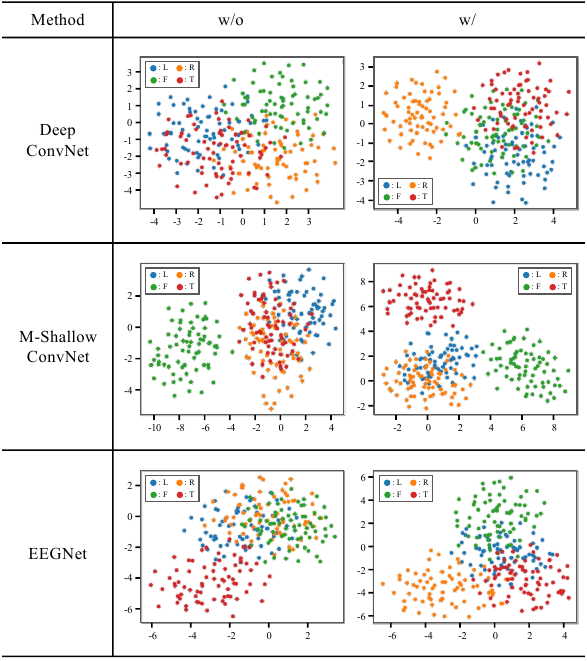}}
\caption{Feature visualization using the \textit{t}--SNE for A5 on BCIC2a. The colors represented the types of each MI task. (w/o: without the proposed method and w/: with the proposed method).}
\label{fig2}
\end{figure}
% %%%%%%%%%%%%%%%%%%%%%%%%%%%%%%%%%%%%%%%%%%%%%%%%%%%%%%%

\section{RESULTS AND DISCUSSION}
\subsection{Evaluation of the Performances}
We used the average accuracy (avg.) and the standard deviation (std.) across all subjects as evaluation metrics. In the evaluation environment, the leave--one--subject--out cross--validation \cite{lee2021subject} was adopted to confirm the classification performances in the cross--subject environment. Additionally, we conducted the ablation study on three conventional models (the DeepConvNet \cite{schirrmeister2017deep}, the M--ShallowConvNet \cite{kim2022rethinking}, and the EEGNet \cite{lawhern2018eegnet}) to demonstrate the effectiveness of the proposed method, as shown in Table I. The DeepConvNet was designed to automatically extract features at various scales by the deep layer structure. The M--ShallowConvnet optimized the layer structure of the ShallowConvNet \cite{schirrmeister2017deep} to alleviate bottlenecks in the training phase. The ShallowConvNet was proposed to capture both temporal and spatial features of EEG signals, which is inspired by the common spatial pattern method. The EEGNet was designed to be robust across various BCI paradigms with the compact structure. The common attribute of these models is the ability to learn features for each frequency band and effectively combine the features extracted from different frequency bands. 

In Table I, the bold text represented high performances when the proposed method was integrated into the conventional models and compared to the results of the classification performances. The DeepConvNet, the M--ShallowConvNet, and the EEGNet showed the performances (avg.(std.)) of 62.35(7.35) \%, 68.67(11.07) \%, and 69.08(6.24) \%, respectively. These results indicated that when the proposed method was integrated into the DeepConvNet and M--ShallowConvNet, the performances (avg.(std.)) were improved by 2.86(0.02) \% and 3.78(0.32) \%, respectively. However, in the case of the EEGNet, we observed a decrease in the performances (avg.(std.)) by 0.11(0.17) \%. We assumed that the EEGNet was designed more compactly compared to other conventional models, resulting in fewer parameters, which could lead to the limited capacity to learn diverse features. Specifically, it suggested that the EEGNet might be insufficient to effectively learn the discriminative features of signals processed by the proposed method compared to raw signals. \\

\subsection{Feature Visualization}
To confirm the effectiveness of the proposed method, we additionally visualized the difference of the feature distributions by the \textit{t}--distributed stochastic neighbor embedding (\textit{t}--SNE), as shown in Fig. 2. We designed the proposed method to be placed at the front of the layers in the conventional models and conducted the comparison of the feature distributions between the cases w/ and w/o our proposed method. The representative subject was selected as A5, who showed common improvement in three conventional models. As an experimental result, the conventional models indicated improved boundary separation for each MI task compared to the performances without the proposed method. Although the EEGNet indicated similar variations in the performances, these results suggested that the feature distributions were effectively clustered for subjects with the improved classification accuracy. However, the feature distributions of some MI tasks continued to overlap in three conventional models. We regarded that solely relying on the encoder based on the conventional models in the cross--subject environment has difficulty in extracting the generalized features. Therefore, we confirmed that raw MI signals contain features with low correlation to MI characteristics, which negatively affect the classification accuracy. Our proposed method can obtain the discriminative feature distributions, attenuating features with low correlation to MI characteristics. Furthermore, the results demonstrated that the proposed method could be robustly integrated with the conventional models in the MI domain.\\

\section{CONCLUSTION}
In this paper, we introduced the end--to--end deep preprocessing method to enhance MI characteristics in EEG--based BCI systems. By leveraging temporal, spatial, graph, and similarity blocks, our proposed method aimed to improve the ability of feature extraction while attenuating features with low correlation to MI characteristics. We used BCIC2a widely adopted in MI domain to evaluate the effectiveness of our proposed method. The results showed the improved performances when our proposed method was integrated into the conventional models. In particular, the EEGNet, which had few differences between the cases w/ and w/o our proposed method, indicated discriminative boundary separation for subjects with the improved performances. However, the encoders based on the conventional models tend to show the low generalization ability for some MI tasks. In future works, we will conduct the experiment using other public datasets to confirm the robustness of our proposed method. Also, we will develop algorithms that can provide more generalized feature extraction from EEG signals related to MI tasks. Therefore, We believe that our study to enhance features of complex brain dynamics has contributed to the progress of BCI systems.\\

\bibliographystyle{IEEEtran}
\bibliography{REFERENCE}

\end{document}